%% file: manuscript.tex
\documentclass[9pt,twocolumn,twoside]{pnas-new}

\templatetype{pnasresearcharticle} 

\newcommand{\pdf}{\mathrm{Pr}}
\newcommand{\GG}{\mathbb{G}}
\newcommand{\MM}{\mathbb{M}}

\title{\mbox{Formulation \& Resolutions of the Red Sky Paradox}}

\author[a]{David Kipping}
\affil[a]{Columbia University, 550 W 120th St., New York, NY 10027}

\leadauthor{Kipping} 

\significancestatement{
Red dwarf stars are the most numerous and long-lived stars in the cosmos, and recent exoplanet discoveries indicate an abundance of rocky, temperate planets around them. This presents an apparent paradox as to why we don’t see a red dwarf in our sky. This “Red Sky paradox” could plausibly be random chance at the 1-in-100 level, but would then come into tension with Copernican Principle. Three additional resolutions to the paradox are outlined, which broadly inhibit the opportunities for complex life to develop around such stars: attenuated emergence rates, truncated evolutionary windows and/or a paucity of suitable habitats. All three appear viable given our present limited knowledge but the potential for future observational tests is explored.
}


\begin{abstract}
Most stars in the Universe are red dwarfs. They outnumber stars like our Sun by a factor of 5 and outlive them by another factor of 20 (population-weighted mean). When combined with recent observations uncovering an abundance of temperate, rocky planets around these diminutive stars, we’re faced with an apparent logical contradiction - why don’t we see a red dwarf in our sky? To address this “Red Sky paradox”, we formulate a Bayesian probability function concerning the odds of finding oneself around a F/G/K-spectral type (Sun-like) star. If the development of intelligent life from prebiotic chemistry is a universally rapid and ensured process, the temporal advantage of red dwarfs dissolves softening the Red Sky paradox, but exacerbating the classic Fermi paradox. Otherwise, we find that humanity appears to be a 1-in-100 outlier. Whilst this could be random chance (resolution I), we outline three other non-mutually exclusive resolutions (II-IV) that broadly act as filters to attenuate the suitability of red dwarfs for complex life. Future observations may be able to provide support for some of these. Notably, if surveys reveal a paucity temperate rocky planets around the smallest (and most numerous) red dwarfs then this would support resolution II. As another example, if future characterization efforts were to find that red dwarf worlds have limited windows for complex life due to stellar evolution, this would support resolution III. Solving this paradox would reveal guidance for the targeting of future remote life sensing experiments and the limits of life in the cosmos.
\end{abstract}

\dates{Competing interest statement: The authors declare no competing interests.}
\doi{\url{www.pnas.org/cgi/doi/10.1073/pnas.XXXXXXXXXX}}

\begin{document}

\maketitle
\thispagestyle{firststyle}
\ifthenelse{\boolean{shortarticle}}{\ifthenelse{\boolean{singlecolumn}}{\abscontentformatted}{\abscontent}}{}

\dropcap{A} basic observable concerning our existence is the star around which
we find ourselves. Although many early thinkers reasoned as much prior to
modern astronomy\footnote{The earliest example appears to be the
pre-Socratic Greek philosopher Anaxagoras circa 450\,BC.}, it wasn't until the
parallax measurement of 61 Cygni by Freidrich Bessel in 1838 \cite{bessel:1838}
that the Sun's banality as being merely another star within the cosmos was
firmly established. Yet despite this, the Sun is in many ways atypical of the
ensemble.

The mass of our Sun ($\equiv 1\,M_{\odot}$), arguably its most
fundamental property, is an order-of-magnitude greater
than the minimum mass ($\simeq 0.08$\,$M_{\odot}$) necessary to generate the
internal conditions required for hydrogen fusion \cite{
burrows:1997,baraffe:1998,chabrier:2000,baraffe:2003,
dieterich:2014,chen:2017}, but two orders-of-magnitude less than the most
massive stars observed (e.g. BI 253; \cite{bestenlehner:2014}). Although the
Sun could thus be reasonably described as ``middleweight'', it is hardly
typical. Much like pebbles on the beach, there are far more small stars than
massive ones, as revealed by studies of the stellar initial mass function
\cite{salpeter:1955,miller:1979,scalo:1986,chabrier:2003}. Indeed,
approximately three-quarters of all stars are classified as M-dwarfs, a range
spanning ${\sim}0.1$-$0.5$\,$M_{\odot}$. Due to their lower masses, the
internal conditions are less intense than that of the Sun and so these stars
have far lower luminosities, up to three orders-of-magnitude less, leading to
cooler surface temperatures. Thus, inhabitants of these stars would see a pale
orange/red disk in their sky, rather than the brilliant
yellow\footnote{Although it should be noted that without atmospheric
interference the Sun appears much closer to white in colour.} disk we see in
ours.

The candle that burns twice as bright burns half as long, as so it is for stars
too. Indeed, the Universe is not yet old enough for any red-dwarfs (aka
M-dwarfs) to have yet exhausted their fuel supply of hydrogen and are expected
to live for ${\sim}$100 billion years for a 0.5\,$M_{\odot}$ star, and a
staggering ${\sim}$10 trillion years for the smallest stars \cite{adams:1997}.
In contrast, stars greater than 1.6\,$M_{\odot}$, which are denoted as spectral
types A-, B- and O-type, pass through their main sequence lifetimes in less
than two billion years \cite{hansen:2004}, as compared to the Sun's 10 billion
year stint. This severely truncates the opportunities for biology to evolve
from simple chemical  systems to complex, self-aware intelligent beings
\cite{malley:2013}. It is thus perhaps no surprise that we do not find
ourselves living around an O/B/A-type star: not only are they intrinsically
rare (comprising less than 1\% of the stellar population) but they simply do
not persist for long enough to foster complex biology \cite{sato:2014}.

Applying this same argument to M-dwarfs, one encounters an apparent
logical contradiction though. All things beings equal, one should expect that
the far longer temporal window of stable luminosity that M-dwarfs enjoy
should yield a greater chance of complexity and intelligence eventually
evolving \cite{spiegel:2012,kipping:2020}. This is compounded by the fact that
M-dwarfs are an order-of-magnitude more abundant than Sun-like
stars. In the same spirit as the Fermi Paradox, we thus find ourselves facing
another apparent logical contradiction dubbed here as the ``red sky
paradox'' - if M-dwarfs are so common and long-lived, why don't we find
ourselves around one?

In the modern era of exoplanet hunting, one might immediately ask whether
M-dwarfs rarely harbor small rocky planets in their habitable zones, and
so perhaps this offers an immediate remedy. Although we discuss this possibility
in much more depth later, we highlight that current population statistics
find that temperate, rocky planets are apparently common around both M-dwarfs
\cite{dressing:2015} and Sun-like stars \cite{bryson:2020}, and so no immediate
observational resolution exists.

The outlined paradox has been previously noted in earlier work
\cite{haqq:2018}. In that work, the authors formulate that the number of
habitable worlds associated with each star as the product of the number of such
stars and the relative widths of their habitable-zones. It is argued here that
this formulation is problematic, since the abundance of planets around each
star type does not appear to be uniform \cite{mulders:2015,hardegree:2019}, and
indeed their properties which may affect habitability (such as composition,
satellite system, impact rate, etc) cannot be reasonably assumed to be uniform
either \cite{shields:2016}. Further, when discussing the longer lives of
M-dwarfs, the authors weight the relative probability of a star being inhabited
by the main-sequence lifetime directly, which is generally incorrect
\cite{spiegel:2012,kipping:2020}. To see why, consider that the probability of
a hypothetical Earth around an FGK star becoming inhabited equals 5\%. If
M-dwarfs live 100 times longer, then by this reasoning the probability of an
Earth around one of these stars becoming inhabited would be 500\% - in other
words, the probability exceeds unity and is improper. In reality, one should
expect the probability to asymptotically approach unity \cite{spiegel:2012,
kipping:2020}.

On this basis, it is worthwhile to revisit this paradox and in this work we
formulate a Bayesian framework to understand the problem, which reveals four
possible resolutions.

\section*{A Bayesian Framework}

The outlined paradox concerns the probability of intelligent
observers emerging on a habitable world around an FGK- versus an M-dwarf star.
If these probabilities are approximately equal, or even favour FGK's, then no
paradox exists. Although we have qualitatively outlined an argument as to why
neither of these are likely true, we will here analytically do so.

Let us denote the probability of intelligent observers emerging on 
a habitable world, given that the world is bound to a FGK-type star, as
$\pdf(I|\GG)$, where ``$I$'' denotes intelligence and ``$\GG$'' is
shorthand for FGK. The specific conditions defining ``habitable''
are not well known and so our definition is operative - they are worlds
with the necessary conditions to yield life and intelligence in infinite
time under stable irradiance. Strictly, we truly do not mean ``infinite''
here, but rather a timescale that greatly exceeds the lifetime of the stars
in question, in order to guard against the more exotic scenario of
Boltzmann brains \cite{linde:2007}.

Following earlier work \cite{carter:2008,spiegel:2012,scharf:2016,chen:2018,
kipping:2020}, we describe the emergence of life and intelligence as a uniform
rate (i.e. Poisson) process, defined by a rate parameter $\lambda_{\GG}$. The
justification, appropriateness and weaknesses of this assumption are described
in detail in the referenced works and we direct the reader to these for that
discussion.

Unlike \cite{kipping:2020}, abiogenesis and intelligence are not treated as
separate and causally dependent processes but rather as a single compound
process that describes the entire process of non-living chemicals developing
into intelligent beings. This essentially absorbs all of the details into the
$\lambda_{\GG}$ term since the numbers of steps and their relationship
to each other is unimportant to the question we seek to address in this work.
The emergence of intelligence has a finite time window within which to occur
($T_{\GG}$), after which the FGK host star leaves the main sequence
leading to the likely terminal extinction of complex life \cite{caldeira:1992}.
Accordingly, one may show that the probability of at least one successful
emergence within this time frame equals (see \cite{spiegel:2012})

\begin{align}
\pdf(I|\GG) &= 1 - \exp(-\lambda_{\GG} T_{\GG}).
\end{align}

Note that the above avoids the pitfalls of producing probabilities greater
than one. In the limit of $T_{\GG} \to 0$ one finds $\pdf(I|\GG) \to 0$,
whereas $T_{\GG} \to \infty$ yields $\pdf(I|\GG) \to 1$, as expected.
Similarly, for M-dwarfs we have

\begin{align}
\pdf(I|\MM) &= 1 - \exp(-\lambda_{\MM} T_{\MM}).
\end{align}

One may now flip this around and consider the probability of finding oneself
around an FGK- or M-dwarf star, given that intelligence emerged. This can
be accomplished through the use of Bayes' theorem, with which one may show that

\begin{align}
\pdf(\GG|I) &= \pdf(I|\GG) \pdf(\GG)/\pdf(I),\nonumber\\
\pdf(\MM|I) &= \pdf(I|\MM) \pdf(\MM)/\pdf(I),
\end{align}

where $\pdf(\GG)$ and $\pdf(\MM)$ represent the prior probabilities of choosing
an FGK-dwarf and M-dwarf; in other words the intrinsic abundance of these stars
in the cosmos, $n_{\GG}$ and $n_{\MM}$. Using a piecewise function, we can now
incorporate these two possibilities as

\begin{equation}
  \pdf(\star|I) = \frac{1}{\pdf(I)} \times
  \begin{cases}
    n_{\GG} (1 - e^{-\lambda_{\GG} T_{\GG}}), & \text{for } \star = \GG \\
    n_{\MM} (1 - e^{-\lambda_{\MM} T_{\MM}}), & \text{for } \star = \MM.
  \end{cases}
\label{eqn:like}
\end{equation}

Note that we have cast the paradox as a dichotomous problem;
intelligence resides around either an FGK-dwarf, else an M-dwarf.
Other seats for life have certainly been speculated about in the
literature, such as brown dwarfs \cite{lingam:2020}, pre-main
sequence stars \cite{ramirez:2014}, evolved stars
\cite{ramirez:2016}, stellar remnants \cite{agol:2011} and rogue
planets \cite{abbot:2011}. However, including these objects, for
which we have much greater uncertainty about their habitable
windows and feasibility, only serves to exacerbate the red sky
paradox. We thus take the conservative approach of neglecting these
for the remainder of this study.

The as yet undefined $\pdf(I)$ term may be simply thought of as a
normalization constant, which can be evaluated through summation
over all possibilities defined to equal unity probability. In doing
so, one finds

\begin{align}
\pdf(I) &= n_{\GG} (1 - e^{-\lambda_{\GG} T_{\GG}}) + n_{\MM} (1 - e^{-\lambda_{\MM} T_{\MM}}).
\label{eqn:norm}
\end{align}

Equipped with a Bayesian formulation of the relevant probabilities, one
can clearly see the various parameters affecting the analysis. Using this,
we describe four possible resolutions to the red sky paradox.

\section*{Resolution I: An Unusual Outcome}

Equipped with Equation~(\ref{eqn:like}), one may now evaluate the
likelihood of finding ourselves around an FGK-dwarf. Just how surprising is
it? Indeed, one possible resolution - dubbed resolution I - is that nothing
is intrinsically different about the emergence of intelligence between
FGK- and M- dwarfs; we're simply an unusual member by finding ourselves
around a yellow dwarf.

All of the terms on the right-hand side of Equation~(\ref{eqn:like})
can be reasonably estimated with the exception of $\lambda_{\GG}$ and
$\lambda_{\MM}$ - the rate at which intelligent life emerges on
habitable worlds around FGK dwarfs. Although we might consider that
these rates are different between the two star types (as we will do
so explicitly in the next section), the simplest assumption is that they are
approximately equal. Thus, we have just one unknown that an be explored
through parameter variation.

For $n_{\GG}$ and $n_{\MM}$, one can simply assign these as the
number of stars of each type. This implicitly assumes that the probability of
finding a habitable world around each type of star is approximately equal.
As discussed earlier, this is broadly consistent with current constraints from
\textit{Kepler} \cite{dressing:2015,bryson:2020}, although very large
uncertainties are in play. The possibility
of this difference being significant is explored in the distinct resolution,
resolution IV, later. With the initial mass function of \cite{kroupa:2001},
the ratio $n_{\MM}/n_{\GG}$ is evaluated to be $4.97$ using the
boundaries $[0.08,0.55]\,M_{\odot}$ for M-dwarfs (M9 to M0 spectral types
\cite{mamajek}) and $[0.55,1.6]\,M_{\odot}$ for FGK-dwarfs (K9 to F0 spectral
\cite{mamajek}).

Assigning singular values for $T_{\MM}$ and $T_{\GG}$ is challenged by the fact
that sub-types within each category have very different lifetimes. One can
calculate the weighted mean time across the various sub-types by weighting with
the relative abundances as computed using the initial mass function and
estimating the main sequence lifetimes as ${\sim}10 (M/M_{\odot})/(L/L_{\odot})$\,Gyr
\cite{hansen:2004}. Doing so, and using the \cite{mamajek} spectral type
values, we find population weighted mean values of $T_{\GG} = 31$\,Gyr and
$T_{\MM} = 651$\,Gyr. This tacitly assumes that complex life is viable
throughout the main-sequence lifetime, although complex life might only
reasonably persist for some fraction of the total \cite{caldeira:1992}. 
Nevertheless, the habitable windows can be reasonably assumed to scale with
the main-sequence lifetimes and thus we emphasize that it is the ratio of
these values that is of greatest import here. Broadly speaking then, M-dwarfs
are ${\sim}5$ times more abundant and ${\sim}20$ times longer lived than
FGK-dwarfs. Together then, this quantifies that the red sky paradox concerns an
imbalance of two orders-of-magnitude.

\begin{figure}
\begin{center}
\includegraphics[width=8.4cm,angle=0,clip=true]{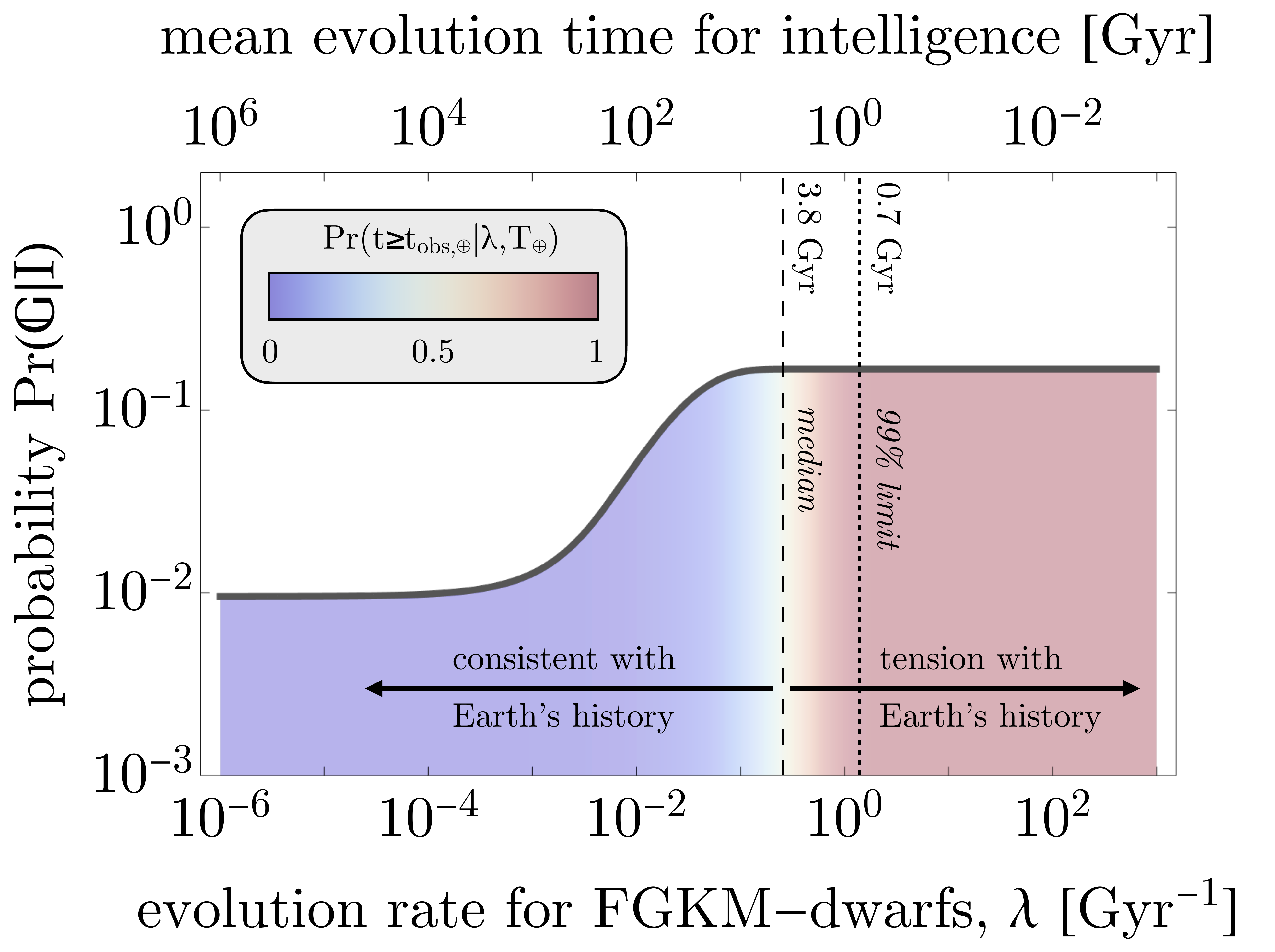}
\caption{
Resolution I to the red sky paradox, where our existence around a FGK-dwarf
is merely an unlikely chance event. The y-axis denote just how unlikely
this is as a function of the rate of intelligence emergence, $\lambda$.
High $\lambda$ values fall into tension with Earth's evolutionary record
(see the SI for more).
}
\label{fig:solution1}
\end{center}
\end{figure}

Using these numbers, we evaluate the likelihood of finding ourselves around
an FGK star using Equation~(\ref{eqn:like}) as a function of the one unknown
parameter, $\lambda$, in Figure~\ref{fig:solution1}. From this, two asymptotes
are revealed in the slow ($\lambda \to 0$) and fast ($\lambda \to
\infty$) emergence rate limits. In the limit of fast intelligence
emergence, corresponding to a cosmos teaming with sentience,
the red sky paradox is largely dissolved, which we dub as resolution I-f
(f for ``fast''), since

\begin{align}
\lim_{\lambda \to \infty} \pdf(\GG|I) &= \frac{n_{\GG}}{n_{\GG}+n_{\MM}} \sim \frac{1}{6}.
\label{eqn:solution1f}
\end{align}

In contrast, in the rare intelligence limit, humanity would need to be a far
more unusual example (which we dub resolution I-s, ``slow''), with

\begin{align}
\lim_{\lambda \to 0} \pdf(\GG|I) &= \frac{n_{\GG} T_{\GG}}{n_{\GG} T_{\GG} + n_{\MM} T_{\MM}} \sim 10^{-2}.
\label{eqn:solution1s}
\end{align}

On the face of it, resolution I-f might seem the most straight-forward
solution then. However, assuming such a fast emergence rate comes into
tension with one hard observable and another softer one. The first hard limit
is the timing of our own arrival in the evolutionary record here on Earth.
Repeating the analysis of \cite{kipping:2020} under a compound process for
abiogenesis + intelligence, as used here, yields a monotonically decreasing
posterior distribution for $\lambda$ peaking at 0 (see the S.I.). From this,
50\% of the posterior parameter space is below $0.26$\,Gyr$^{-1}$ indicating
that very low $\lambda$ rates are fully compatible with the timing of our own
arrival in Earth's evolutionary record \cite{carter:2008}. Although the 99\%
upper limit permits faster rates, this inference ignores any biological
constraints on the rate of emergence and thus may be unrealistically expedient.
Further, the softer constraint is that fast $\lambda$ values exacerbate the
classic Fermi paradox, since it leads to a cosmos teaming with
intelligence that eludes detection. On this basis, there are good reasons to
be skeptical that a fast emergence rate naturally explains the red sky paradox
via resolution I-f.

Turning back to resolution I-s, then, this is also hardly satisfying by
simply stating we are a 1-in-100 outlier. Whilst it is indeed technically
possible, it comes into tension with the Copernican principle that posits that
our place in the Universe is typical, and is indeed often treated as a basic
assumption in our studies of the cosmos \cite{clarkson:2008}. We thus consider
other resolutions in what follows.

\section*{Resolution II: Inhibited Life Under a Red Sky}

The paradox can be resolved if $\pdf(\GG|I) \gtrsim \pdf(\MM|I)$. Whilst one
can be more specific than this and assign various confidence intervals
(e.g. 95\%), the simplicity of our model and approach does not warrant such an
analysis, in our view. Applying this condition to Equation~(\ref{eqn:like}),
and re-arranging to make $\lambda_{\MM}$ the subject, one finds

\begin{align}
\lambda_{\MM} &\lesssim - T_{\MM}^{-1} \log\big( 1 - (n_{\GG}/n_{\MM}) (1 - e^{-\lambda_{\GG}} T_{\GG}) \big).
\label{eqn:solution2}
\end{align}

\begin{figure}
\begin{center}
\includegraphics[width=8.4cm,angle=0,clip=true]{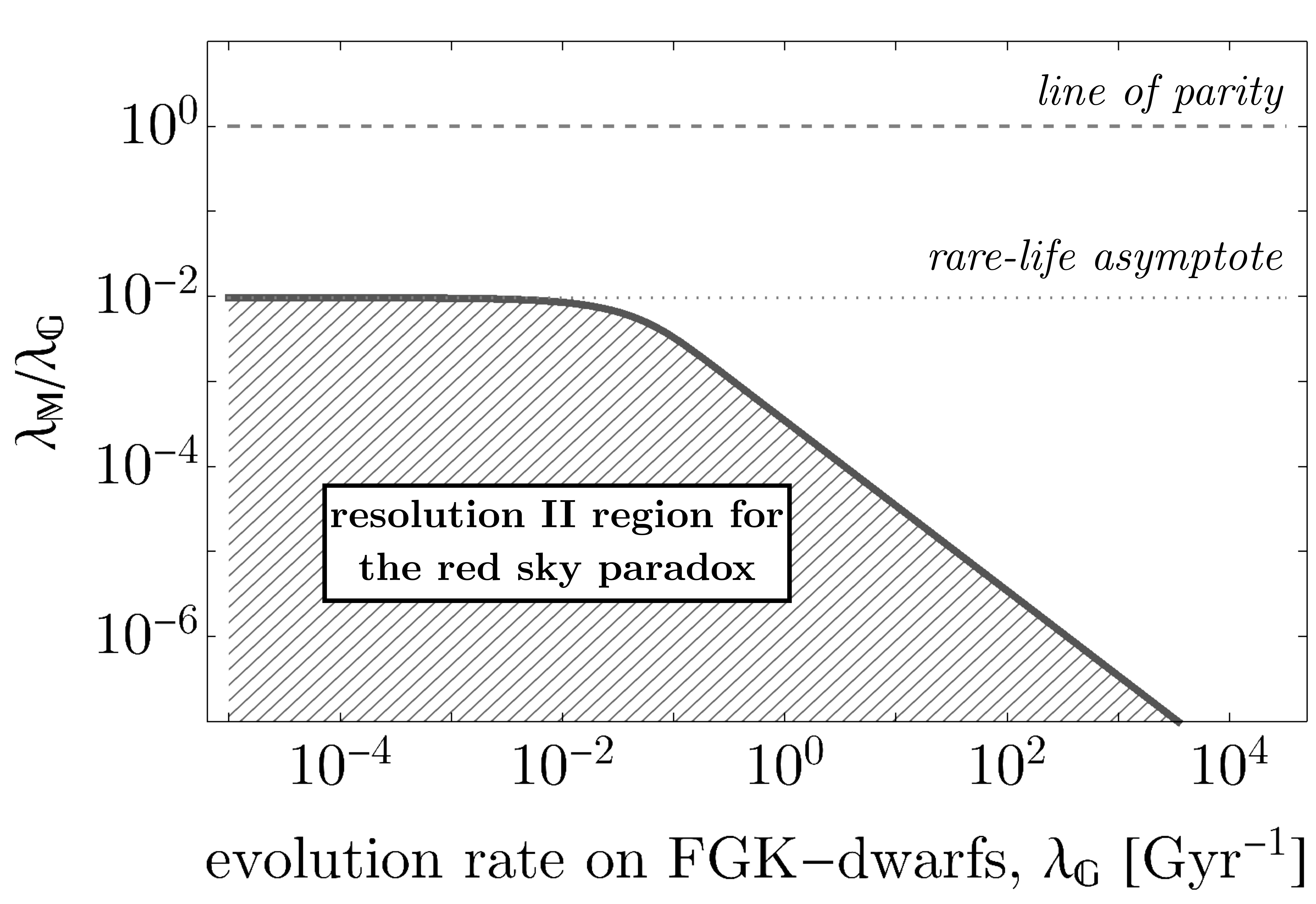}
\caption{
Resolution II to the red sky paradox, where the emergence rate
of intelligent life, $\lambda$, is much slower on
M-dwarfs than FGK-dwarfs. The hatched region shows the zone
necessary to resolve the paradox such that $\pdf(\GG|I) \gtrsim
\pdf(\MM|I)$, which is at least two orders-of-magnitude below
the line of parity.
}
\label{fig:solution2}
\end{center}
\end{figure}

Although $\lambda_{\GG}$ is broadly unknown (although see
\cite{kipping:2020}), one can evaluate Equation~(\ref{eqn:solution2}) as a
function of this unknown as done earlier. Since we are primarily interested
in the relative capability of each star type producing intelligent life, the
ratio $\lambda_{\MM}/\lambda_{\GG}$ is of particular interest. This function is
shown in Figure~\ref{fig:solution2}, where one can see that for small
$\lambda_{\GG}$ values (a rare intelligence universe),
$\lambda_{\MM}/\lambda_{\GG}$ plateaus at around 1\%. A turn-over begins
at around $\lambda_{\GG} = 10^{-1.5}$\,Gyr$^{-1}$, which corresponds to
even odds of each habitable planet around a FGK-dwarf spawning
intelligence. Beyond this point, one requires $\lambda_{\MM} \ll
\lambda_{\GG}$, which is because at high $\lambda_G$, even the
shorter lived FGK-stars become widely inhabited and thus $\lambda_M$ has
to dive down rapidly to prevent the more numerous, longer-lived M-dwarfs
dominating. One can thus
see that this resolution may be analytically expressed as requiring

\begin{align}
\frac{\lambda_{\MM}}{\lambda_{\GG}} \lesssim
\frac{T_{\GG} n_{\GG}}{T_{\MM} n_{\MM}} \sim 10^{-2}.
\end{align}

In other words, the probability of intelligent life emerging on M-dwarfs
would need to be at least two-orders of magnitude less than than that of
FGK-dwarfs. Certainly, much theoretical work has questioned the plausibility
of complex life on M-dwarfs \cite{shields:2016},
with concerns raised regarding tidal locking and atmospheric collapse
\cite{joshi:1997,wordsworth:2015,heng:2020}, increased exposure to the effects
of stellar activity \cite{segura:2010,rugheimer:2015a,rugheimer:2015b},
extended pre-main sequence phases \cite{baraffe:1998,baraffe:2015},
and the paucity of potentially beneficial Jupiter-sized companions
\cite{horner:2008,horner:2009,johnson:2012}. On this basis, there is good
theoretical reasoning to support resolution II, although we
emphasize that it remains observationally unverified.

\section*{Resolution III: A Truncated Window for Complex Life}

Another way to inhibit life on M-dwarfs is not to attenuate $\lambda_{\MM}$,
but instead truncate the time window available, $T_{\MM}$. Terrestrial
worlds forming in the main-sequence habitable-zones of
M-dwarfs will be subject to an initial phase of high irradiance during the
${\sim}$Gyr pre-main sequence phase \cite{baraffe:2015}, potentially pushing
them into a runaway greenhouse state \cite{kasting:1988} that persists
thereafter \cite{luger:2015}. Although one might discount such worlds, their
more distant orbiting siblings may enjoy a brief episode of habitability during
this initial ${\sim}$Gyr phase \cite{ramirez:2014}.

As before, we proceed by setting $\pdf(\GG|I) \gtrsim \pdf(\MM|I)$ but now
instead solve for $T_{\MM}$. Under this resolution, which is distinct from
resolution I, there is no significant difference between $\lambda_{\MM}$
and $\lambda_{\GG}$ and thus both are set to a universal value of $\lambda$,
to give

\begin{align}
T_{\MM} &\lesssim -\lambda^{-1} \log\big( 1 - (n_{\GG}/n_{\MM}) (1-e^{-\lambda T_{\GG}}) \big).
\label{eqn:solution3}
\end{align}

\begin{figure}
\begin{center}
\includegraphics[width=8.4cm,angle=0,clip=true]{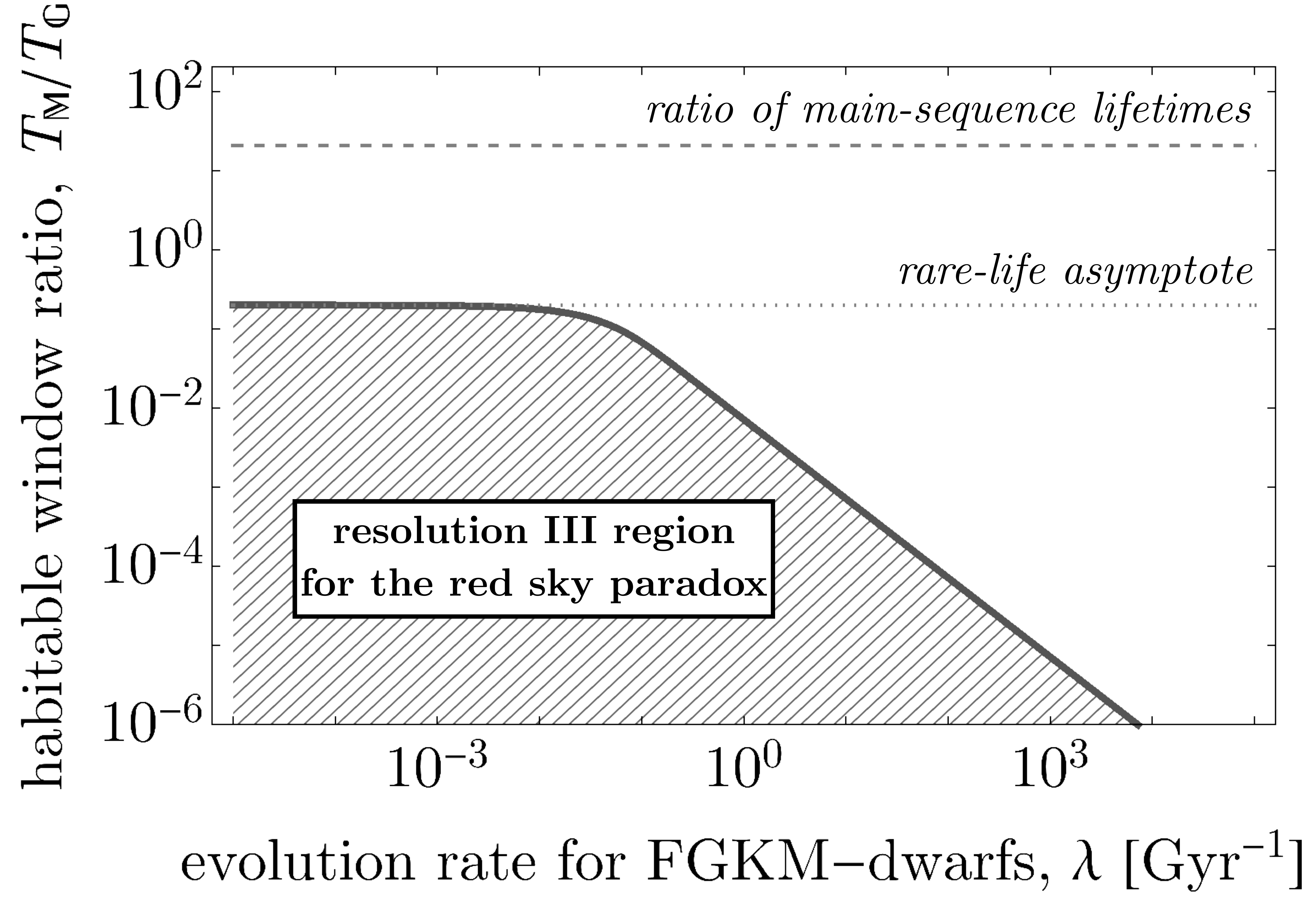}
\caption{
Resolution III to the red sky paradox, where the temporal window for
evolution producing complex life, $T$, is less for
M-dwarfs than FGK-dwarfs. The hatched region shows the zone
necessary to resolve the paradox such that $\pdf(\GG|I) \gtrsim
\pdf(\MM|I)$, which is at least five below parity, and two
orders-of-magnitude below the expected value by comparing the
population-weighted main-sequence lifetimes.
}
\label{fig:solution3}
\end{center}
\end{figure}

Equation~(\ref{eqn:solution3}) displays similar functional behaviour to
Equation~(\ref{eqn:solution2}), as can be seen in Figure~\ref{fig:solution3}.
Here, we find the plateau occurs at approximately one-fifth of $T_{\GG}$
when using our canonical values, or expressing analytically we have

\begin{align}
\frac{T_{\MM}}{T_{\GG}} \lesssim
\frac{n_{\GG}}{n_{\MM}} \sim \frac{1}{5}.
\end{align}

Thus, resolution III requires that the habitable window of M-dwarfs is
${\gtrsim}5$ times less than that of FGK-dwarfs. Such a value aligns
with the pre-main sequence lifetimes ranging from 200 \,Myr (M1) to 2.5\,Gyr
(M8), which is indeed ${\lesssim}5$ times the main-sequence lifetimes of
FGK-stars, and thus we consider this a viable explanation.

\section*{Resolution IV: A Paucity of Pale Red Dots}

The final way one can manipulate Equation~(\ref{eqn:like}) to resolve the
paradox is via the number of seats for life. Thus far, we have tacitly
assumed that the occurrence rate of habitable worlds around FGK-dwarfs
is approximately the same as that as M-dwarfs. Recall that our definition
of a habitable world is one that will eventually culminate in complex,
intelligent life in infinite time under stable irradiance. This is not
an observable property, but modern astronomy is able to probe the
occurrence rate of approximately Earth-sized planets in the temperate
regions around stars where liquid water could be stable on their
surfaces.

Statistical analysis of the \textit{Kepler} exoplanet population reveals that
$16_{-7}^{+17}$\% of the observed M-dwarfs host Earth-sized planets in a
conservatively defined temperate zone \cite{dressing:2015}, with other studies
finding compatible values \cite{hardegree:2019,tuomi:2019}. The situation for
\textit{Kepler}'s FGK-stars is less clear with significant disagreement between
different studies. For example, values of $2.8_{-1.9}^{+4.7}$\%,
$1.9_{-0.8}^{+1.0}$\% and $1.3_{-0.6}^{+0.9}$\% have been reported
\cite{catan:2011,dfm:2014,bryson:2019}, but so too have values as high as
$103_{-10}^{+10}$\% and $124_{-5}^{+6}$\% \cite{traub:2012,garrett:2018}. The
most recent analysis lands somewhere in the middle at $37_{-21}^{+48}$\%
\cite{bryson:2020}. At the present time then, there is no clear evidence that
the occurrence rate of temperate, Earth-sized planets is distinct between
these stars.

However, there are crucial ways in which this could be wrong. First, the
smallest M-dwarfs hardly feature in the \textit{Kepler} catalog despite
comprising the majority of M-dwarfs. This is due to Malmquist bias
\cite{malmquist:1922} - the smallest stars produce insufficient luminosities
to be detected in such surveys in large numbers. It is deeply unclear whether
the exoplanet population of M0-M2 dwarfs is representative of the entire
M-dwarf sample and thus it's possible that this could either resolve, or
exacerbate, the red sky paradox. Second, the occurrence rate of temperate,
Earth-sized planets is often presumed to be a proxy for the occurrence rate
of habitable worlds, but this too could be challenged. Moons are completely
ignored in this calculus, as are the subtle effects of internal composition,
atmospheric composition, obliquity, rotation and circumstellar environment.
For example, the detected population of Earth-sized planets around M-dwarfs
could be dominated by photoevaporated cores of sub-Neptunes
\cite{carrera:2018}. Accordingly, it is quite plausible that there are
substantial differences in the frequency of habitable abodes between the two
categories.

To account for this, let us modify $n_{\GG} \to \epsilon_{\GG} n_{\GG}$ and
$n_{\MM} \to \epsilon_{\MM} n_{\MM}$ in Equation~(\ref{eqn:like}), where
$\epsilon$ represents the fraction of stars with one or more habitable worlds
around them. Following this through and setting $\pdf(\GG|I) \gtrsim
\pdf(\MM|I)$ as before, we can solve for $\epsilon_{\MM}/\epsilon_{\GG}$ to be

\begin{align}
\frac{\epsilon_{\MM}}{\epsilon_{\GG}} &\lesssim \frac{n_{\GG}}{n_{\MM}} \Big( \frac{1-e^{-\lambda T_{\GG}}}{1-e^{-\lambda T_{\MM}}} \Big)
\label{eqn:solution4}
\end{align}

\begin{figure}
\begin{center}
\includegraphics[width=8.4cm,angle=0,clip=true]{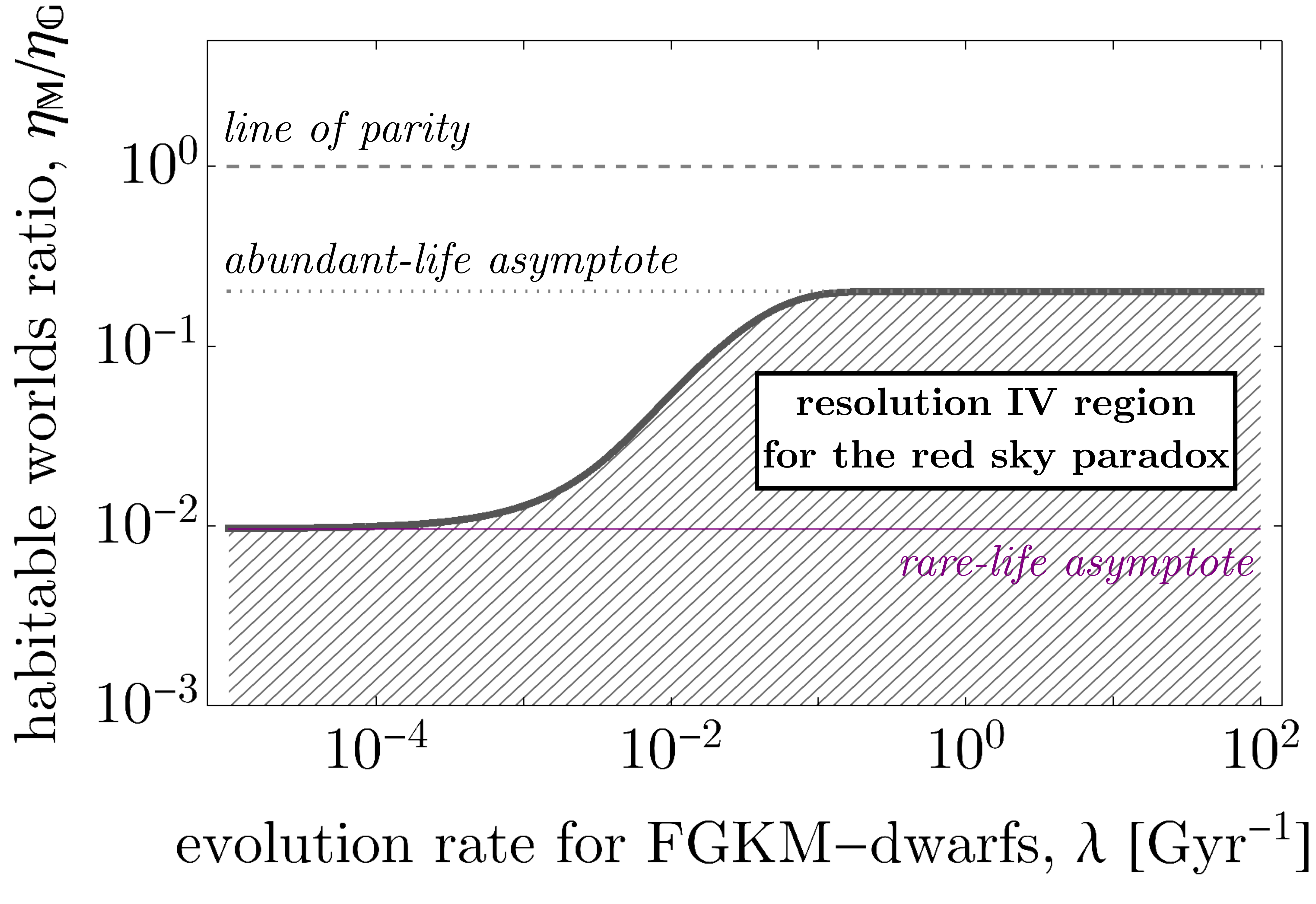}
\caption{
Resolution IV to the red sky paradox, where the occurrence rate of habitable
worlds bound to M-dwarfs is less than FGK-dwarfs. The hatched region shows the
zone necessary to resolve the paradox such that $\pdf(\GG|I) \gtrsim
\pdf(\MM|I)$.
}
\label{fig:solution4}
\end{center}
\end{figure}

where we have again assumed that the emergence rate is approximately
the same for habitable worlds bound to M-dwarfs and FGK-dwarfs. The
functional form of Equation~(\ref{eqn:solution4}) is plotted in
Figure~\ref{fig:solution4}, where the existence of two asymptotes becomes
apparent. As with the previous resolutions, we again find a rare-life asymptote
as $\lambda \to 0$ which tends to

\begin{align}
\lim_{\lambda \to 0} \frac{\epsilon_{\MM}}{\epsilon_{\GG}} &\lesssim \frac{n_{\GG} T_{\GG}}{n_{\MM} T_{\MM}} \sim 10^{-2}.
\label{eqn:solution4s}
\end{align}

In this case, intelligent life is rare amongst the cosmos and spawns
universally between M- and FGK-dwarfs, but habitable worlds are at least
two-orders of magnitude less common around M-dwarfs than FGKs. Let us denote
that as resolution IV-s. Two orders-of-magnitude is a considerable difference
making this a particularly interesting explanation. This would require that the
vast majority of many known Earth-sized, temperate planets around M-dwarfs
\cite{dressing:2015} are somehow inhospitable to life, or that the late-type
M-dwarfs (low mass end) rarely host habitable worlds.

The other asymptote occurs as $\lambda$ grows large, corresponding to
intelligent-life spawning everywhere. Here, resolution IV-f, can be described
as

\begin{align}
\lim_{\lambda \to \infty} \frac{\epsilon_{\MM}}{\epsilon_{\GG}} &\lesssim \frac{n_{\GG}}{n_{\MM}} \sim \frac{1}{5}.
\label{eqn:solution4f}
\end{align}

The requirement imposed by resolution IV-f is more palatable in the sense of
necessitating a meager factor of 5 difference in habitable world occurrence
rates. However, resolution IV-f is also somewhat dubious for the same
reasons discussed earlier with I-f. On this basis, we tend to favour IV-s over
IV-f, at the present time.

\section*{Conclusions}

In this work, it has been argued that our emergence around an FGK-dwarf star is
ostensibly in tension with the fact that smaller M-dwarf stars dominate the
stellar population and live much longer lives. If M-dwarfs are so numerous
(in space and time), why don't we live around one? In many ways, the paradox
is conceptually analogous to the Fermi paradox - if life is presumed to be
common, why don't we see evidence for alien life anywhere?

Like the Fermi paradox, a straight-forward solution is that we are simply
outliers and our observation only appears in tension with the logical argument
because we are a tail-end member of the distribution. Dubbed resolution I in
this work, we find that if intelligent life emerges rapidly, the large
temporal advantage that M-dwarfs enjoy is dissolved and thus the
``surprisingness'' of our yellow host star is modest, given by the ratio of the
abundance of M-dwarfs versus FGKs. However, this solution exacerbates
the related Fermi paradox and even starts to come into tension with the
evolutionary record observed on Earth \cite{kipping:2020}. In the rare
intelligence scenario, tension on the Fermi Paradox is relaxed but tension
on the red sky paradox exacerbates, making our existence a ${\sim}1$\% outlier.
Although we could simply accept our existence as unusual, this is inconsistent
with the Copernican Principle and is hardly a satisfying resolution.

Three other resolutions are proposed, by altering the various terms governing
the likelihood function. Equipped only with our present and limited constraints
on these terms, all of them are ostensibly viable. If resolution I is rejected,
then one or more of these three must hold true: II) the emergence rate of
intelligence is slower for M-dwarfs, III) the available time for intelligence
emergence is truncated for M-dwarfs, and/or IV) M-dwarfs have fewer habitable
worlds.

It is possible that resolution IV could find observational support in
the near term. The occurrence rate of Earth-sized,
temperate planets around late-type M-dwarfs is not well known, but if it could
be established to be much smaller than early-Ms and FGKs, this would provide
support for resolution IV. In such a case, this may actually be good news for
astrobiologists, because it permits the emergence of intelligence on the
early M-dwarfs subset. Since we know of a population of such worlds already
\cite{dressing:2015}, one could maintain justified optimism concerning future
efforts to remotely detect life on these worlds.

If Earths are found to be common around late M-dwarfs, it will not
be possible to further test resolution IV until we can assess if planets
are truly capable of harboring complex life from remote observations. In this
scenario, it is still possible that resolution IV operates, but in a more
nuanced manner than the simple prevalence of planets. For example, such worlds
may have less stable climates/atmospheres as a result of tidal locking
\cite{joshi:1997,wordsworth:2015,heng:2020}. If not, resolutions II
\& III become increasingly favourable.

Resolution III could gain observational support if Earth-sized planets in the
habitable-zone of M-dwarfs are consistently demonstrated to be runaway
greenhouses, something potentially testable with JWST
\cite{barstow:2016} and hypothesized by \cite{luger:2015}. Although reasoned
speculation can be considered regarding resolution II, as a
direct statement about exolife's evolutionary development it would likely be
untestable with any conceived missions and may perhaps only gain through
support by deductive elimination of the stated alternatives. Ultimately,
resolving the red sky paradox is of of central interest to astrobiology and
SETI, with implication as to which stars to dedicate our resources to, as well
as asking a fundamental question about the nature and limits of life in the
cosmos.

Data Availability Statement: All data used in this work is fully stated in the
text of this paper.

\acknow{DK thanks Methven Forbes, Tom Widdowson, Mark Sloan, Laura Sanborn, Douglas Daughaday, Andrew Jones, Marc Lijoi, Elena West, Tristan Zajonc, Chuck Wolfred, Lasse Skov, Alex de Vaal, Jason Patrick-Saunders, Stephen Lee, Zachary Danielson, Vasilen Alexandrov, Chad Souter, Marcus Gillette, Tina Jeffcoat, Jason Rockett, Scott Hannum \& Tom Donkin.}

\showacknow{} 

\bibliography{pnas-sample}

\newpage

\onecolumn

\begin{center}
{\Huge \textbf{Supplementary Information}}
\end{center}

\input{SI.tex}

\end{document}

%% file: SI.tex
\section*{Posterior Probability Distribution of $\lambda$}

Here, we derive the posterior probability distribution for the
emergence rate of life and intelligence, as conditioned upon the
timing of the emergence of intelligent life on Earth. Our approach
closely follows that of \cite{kipping:2020}, with the main difference
being that this work considers abiogenesis and evolution to intelligence
as a single compound process, rather than attempting inference on the
two separate stages as was done in \cite{kipping:2020}.

We start by noting that the probability distribution for the time, $t$,
to obtain a ``success'' (which here equates to an intelligence emerging) from a
Poisson process of rate $\lambda$ is an exponential characterized by

\begin{align}
\pdf(t|\lambda) &= \lambda e^{-\lambda t}.
\end{align}	

Constraining that the success must occur within a time-frame $T$, and
normalizing appropriately, this becomes

\begin{equation}
\pdf(t|\lambda)= \frac{1}{\pdf(I)} \times
  \begin{cases}
    \frac{\lambda e^{-\lambda t}}{1-e^{-\lambda T}}, & \text{for } t < T \\
    0, & \text{otherwise }.
  \end{cases}
\label{eqn:1}
\end{equation}

In the case of the Earth, and again using the values of \cite{kipping:2020},
the time for intelligence to emerge was approximately $t_{\mathrm{obs}} =
4.4$\,Gyr. There is some uncertainty about how quickly life began, but the
uncertainty is considerably less than the scale of this value. Given the
approximate nature of our model, and the very coarse constraints that will
ultimately culminate from this analysis, this does not significantly influence
what follows.

Accordingly, setting $t \to t_{\mathrm{obs}}$ in Equation~({\ref{eqn:1})
defines a Bayesian likelihood function where the ``data'' is
$t_{\mathrm{obs}}$. To infer the posterior distribution of $\lambda$,
we require an appropriate prior for $\lambda$. For reasons discussed at
length in \cite{kipping:2020}, the objective Bernoulli prior is best
suited for this problem, which has the form

\begin{align}
\pdf(\lambda) &= \frac{T}{\pi\sqrt{e^{\lambda T}-1}}.
\end{align}	

Combining this with the likelihood function yields a unnormalized
posterior of

\begin{align}
\pdf(\lambda|t_{\mathrm{obs}}) &\propto \frac{ \lambda e^{\lambda(T-t_{\mathrm{obs}})} }{ e^{\lambda T} - 1 }.
\end{align}

By integration, the normalization constant is the reciprocal of
$T^{-2} \zeta[2,t_{\mathrm{obs}}/T]$. The posterior distribution
is plotted in Figure~\ref{fig:SI}, where we mark the median and
99\% quantile.

\begin{figure}[b!]
\begin{center}
\includegraphics[width=12.0cm,angle=0,clip=true]{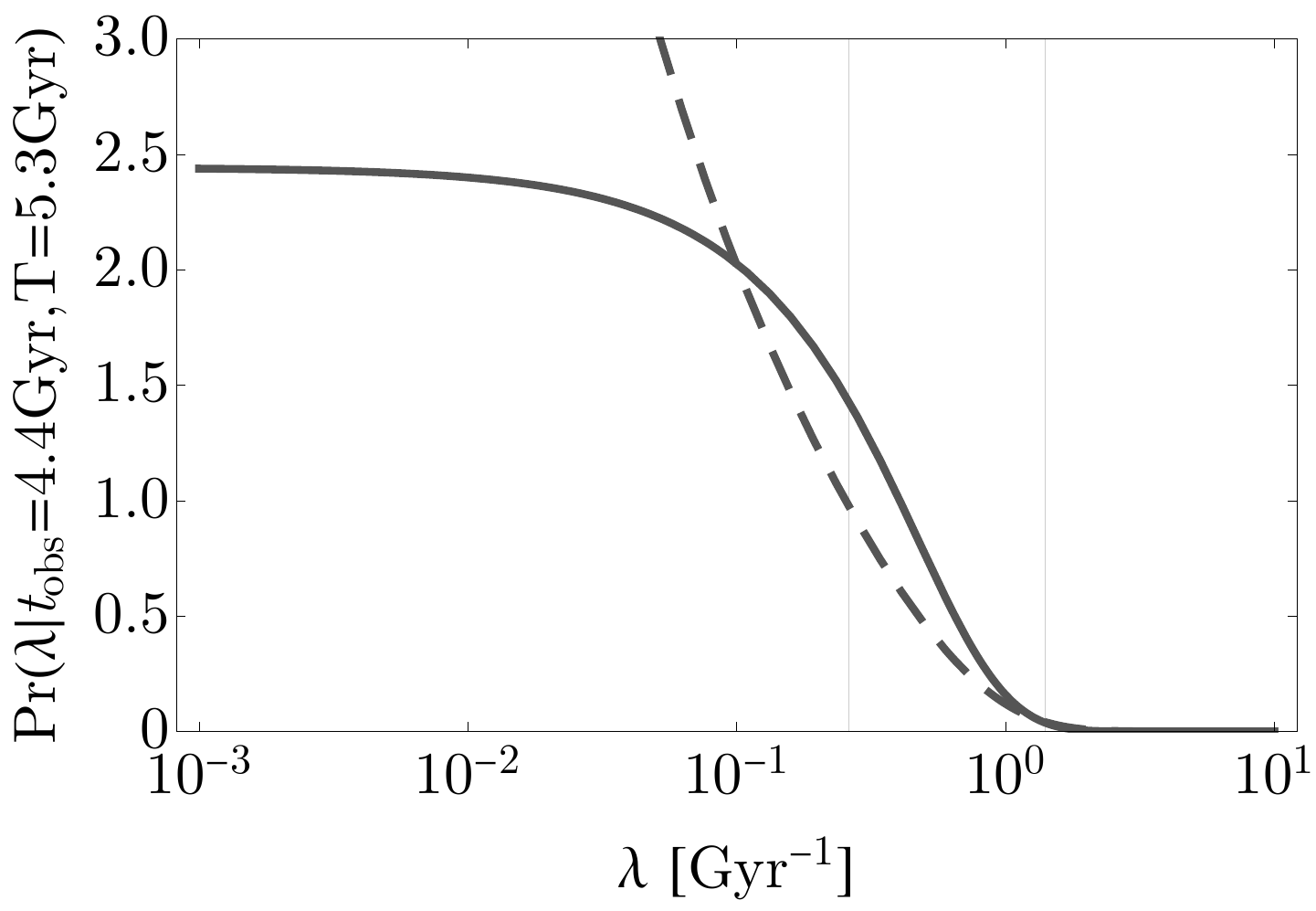}
\caption{
Posterior probability distribution of $\lambda$ (solid) when conditioned upon
the timing of Earth's evolutionary record. The prior is denoted by the curved
dashed line. The left-most vertical dotted line represents the 50\% quantile,
whereas the right-most is the 99\% quantile.
}
\label{fig:SI}
\end{center}
\end{figure}
